\documentclass[aps,prd,superscriptaddress,notitlepage,11pt,final]{revtex4-1}

\usepackage[utf8]{inputenc}  
\usepackage[T1]{fontenc}

\usepackage[pdftex]{graphicx}
\usepackage[pdftex]{xcolor}
\usepackage[pdftex]{hyperref}

\usepackage{amsmath}
\usepackage{amssymb}
\usepackage{amsfonts}
\usepackage{bookmark}
\usepackage{color}
\usepackage{bm}
\usepackage{mathtools}
\usepackage{booktabs}
\usepackage{bbm}
\usepackage{comment}
\usepackage{appendix}
\usepackage{subcaption}

\graphicspath{{./figure_arxiv/}}

\begin{document}
\title{PINNs Study for the Bekki-Nozaki Chaos in the 
Non-linear Schr\"{o}dinger equation}

\author{N. Sawado~}
\email{sawadoph@rs.tus.ac.jp}
\affiliation{Department of Physics and Astronomy, Tokyo University of Science, Noda, Chiba 278-8510, Japan}

\author{Y. Shimazaki~}
\email{shimazakitus@gmail.com}
\affiliation{Department of Physics and Astronomy, Tokyo University of Science, Noda, Chiba 278-8510, Japan}

\author{Y. Suzuki~}
\email{ytszkyuta@gmail.com}
\affiliation{Department of Physics and Astronomy, Tokyo University of Science, Noda, Chiba 278-8510, Japan}

\begin{abstract}
In this paper we study chaotic behavior in the forced dissipative non-linear Schr\"{o}dinger equation, 
so called the Bekki-Nozaki equation.
Chaotic systems are often seen in a strong sensitivity to initial conditions, 
leading to error accumulation over time when traditional numerical methods are applied.
To address this difficulty, we employ Physics-Informed Neural Networks(PINNs), a mesh-free deep learning framework.
PINNs mitigate error accumulation in chaotic systems by solving partial differential equations without discretizing the computational domain.
We demonstrate that PINNs successfully reproduce chaotic behavior of the Bekki-Nozaki equation.
The results of the inverse analysis indicate a correlation between the governing equation's predictability 
and its chaotic nature of the solution.

\end{abstract}

\maketitle

\section{Introduction\label{sec:1intro}}

In non-linear wave systems, those that admit an infinite number of analytical solutions and conserved quantities are referred to as integrable systems.
The partial differential equations(PDEs) governing integrable systems often possess soliton solutions.
PDE systems with only a finite number of conserved quantities, sometimes called quasi-integrable systems, can also exhibit soliton-like structures, such as solitary waves or vortex solutions.
Many integrable systems have been extensively studied in the literature.
In contrast, quasi-integrable systems have received less attention, and their analysis often presents challenges, particularly regarding long-term behavior and multi-soliton collision processes.
In quasi-integrable systems, chaotic properties emerge in the time evolution of the solutions.

In the present paper, we study chaos in the non-linear Schr\"{o}dinger(NLS) equation with some forced perturbative terms, 
so called the Bekki-Nozaki equation~\cite{BekkiNozaki1983,BekkiNozaki1984}:
\begin{align}
iq_t + q_{xx} + 2|q|^2q
= i\varepsilon_1 \exp(i\omega t) 
+ i\varepsilon_2 \exp(2i\omega t)
+ i\gamma q_{xx}
\quad q:= q(x,t),
\label{eq:Bekki-Nozaki_sec1}
\end{align}
where $\varepsilon_1, \varepsilon_2$ and $\gamma(>0)$ are assumed to be small real constants, 
and the fundamental frequency $\omega$ of the external forces is chosen to be positive.
By carefully tuning these parameters, solutions with an attractor or a limit cycle can be obtained.
Analysis by traditional PDE solvers premise discretizing mesh of the computational domain and 
inevitably suffers from various types of error accumulation over time.
In chaotic systems, such errors could have incorrect results due to their sensitivity to initial conditions.
A mesh-independent algorithm is therefore an efficient tool for studying chaotic PDEs.
We apply a deep learning approach called Physics-Informed Neural Networks(PINNs) to analyze~\eqref{eq:Bekki-Nozaki_sec1},  
which was originally introduced by Raissi et al.~\cite{RaissarxivI}.
PINNs incorporate physical laws into conventional neural networks, allowing the solution of PDEs without discretizing the computational domain in space or time.
There are numerous studies of PINNs for the integrable equations as the Burgers equation~\cite{RaissarxivI,RaissarxivII}, 
the KdV equation~\cite{JAGTAP2020113028} and also of the 
NLS equation~\cite{ZHOU2023164,PENG2024}. 
Also, analysis of the quasi-integrable equations in terms of PINNs are studied
~\cite{Nakamula:2024cmx,Nakamula:2024qtp}, 
where several nontrivial results of the inverse analysis are discussed.  
Both forward and inverse PINNs are examined in order to investigate the chaotic property of~\eqref{eq:Bekki-Nozaki_sec1}.

\section{The Bekki-Nozaki Chaos\label{sec:2bekkinozaki}}

\begin{figure}[t]
\begin{tabular}{cc}
\begin{minipage}{.46\textwidth}
\centering
\includegraphics[width=1.10\linewidth]{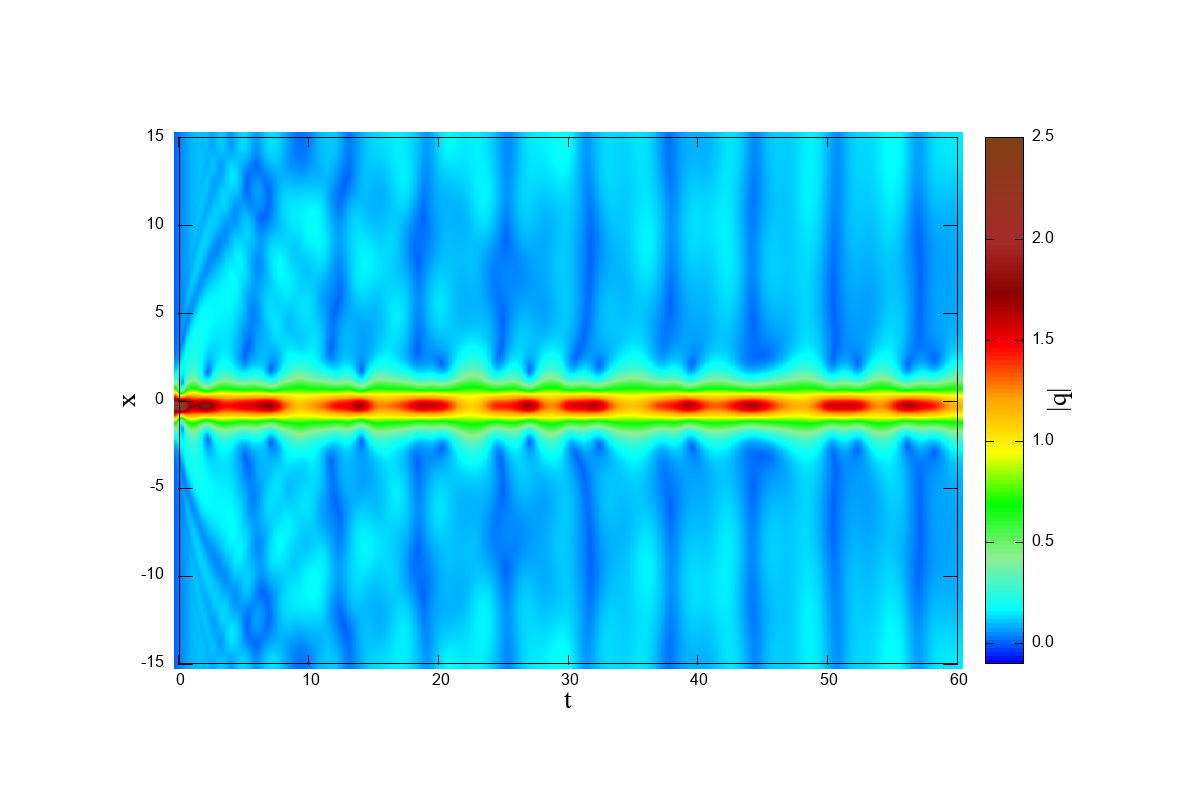}
\subcaption{\label{fig:contour_sec2}
The contour plot}
\end{minipage}
\hspace{1.0pc}
\begin{minipage}{.46\textwidth}
\centering
\vspace{0.5cm}
\includegraphics[width=0.90\linewidth]{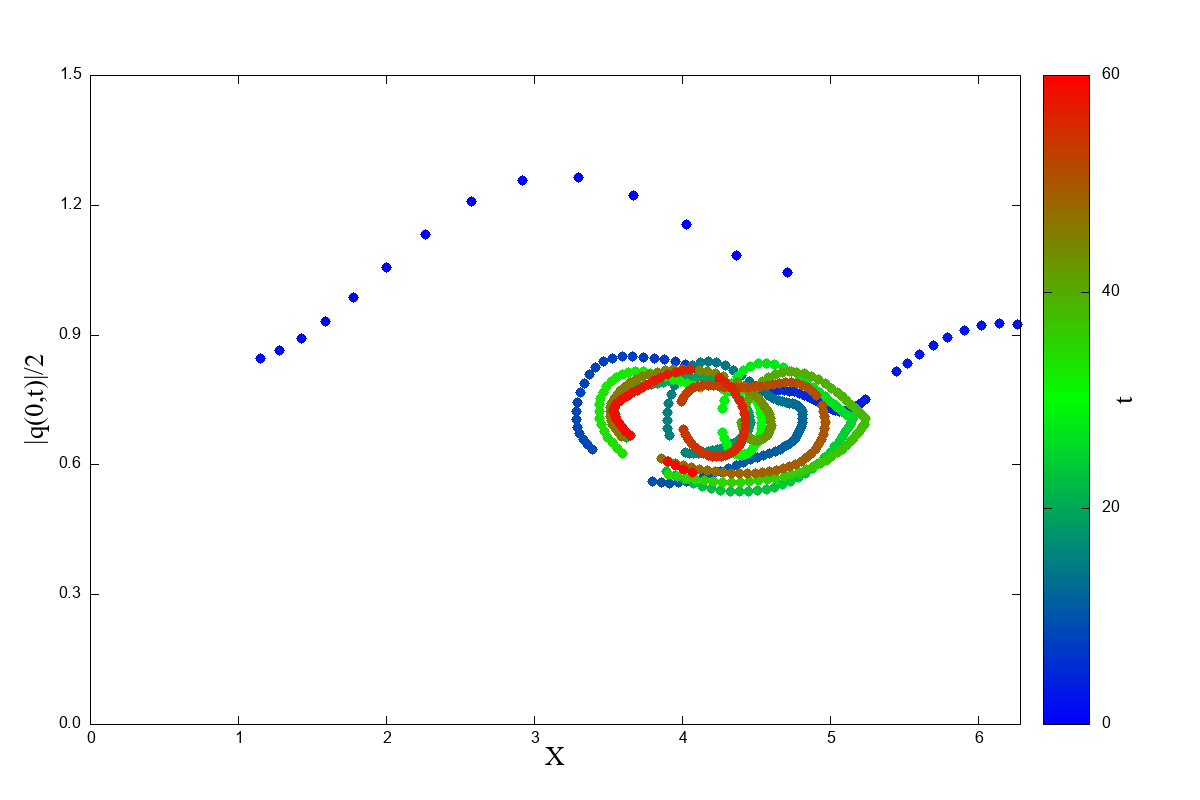}
\vspace{0.5cm}
\subcaption{\label{fig:attractor_sec2}
The trajectory}
\end{minipage}
\end{tabular}
\caption{\label{fig:solution1_sec2}
The behavior of the solution of Eq.~\eqref{eq:Bekki-Nozaki_sec1} with the parameters $A=1.2,\eta_0=0.871$,
$\varepsilon_1=\varepsilon_2=0.0405,\gamma=0.05$ and $\omega=1$.
(a) The contour plot of $|q(x,t)|$ versus $x$ and $t$.
(b) The trajectory of the attractor in the phase space $(\chi(t), |q(x_m,t)|/2)$.}
\end{figure}

We present our numerical results of \eqref{eq:Bekki-Nozaki_sec1} as demonstrated in ~\cite{BekkiNozaki1984}. 
The initial condition is a static one-soliton solution
\begin{align}
q(x,0) = 2A\eta_0 \text{sech}\left(2\eta_0 x\right)\,.
\label{eq:Bekki-Nozaki_ic_sec2}
\end{align}
In Fig.\ref{fig:solution1_sec2}(a), we present the contour plot of $|q(x,t)|$. 
Fig.\ref{fig:solution1_sec2}(b) is
the trajectory of the attractor plot in the phase space $(\chi(t),|q(x_m,t)|/2)$, 
where $\chi(t)=2\omega t-\mathrm{arg}\left\{q(x_m,t)\right\}-\pi/2$ and $x_m(t)$ is the position where the norm $|q(x,t)|$ 
has its maximal value. 
The parameters of the initial condition is $A=1.2,\eta_0=0.871$, 
and for the perturbative terms, we set $\varepsilon_1=\varepsilon_2=0.0405,\gamma=0.05$ and $\omega=1.0$.
The numerical integration of \eqref{eq:Bekki-Nozaki_sec1} is performed in terms of the fourth-order Runge-Kutta method.
The analytical solutions of the integrable NLS equation can be written as
\begin{align}
q(x,t) = \text{sech}(x)\exp(it)
\label{eq:bright-soliton_sec2}
\end{align}
which is known as the bright soliton.
When small dissipation and external forces are applied to the integrable NLS equation, 
the soliton~\eqref{eq:bright-soliton_sec2} can continue to function as the attractor of the system
~\cite{BekkiNozaki1983,BekkiNozaki1984}.

\begin{figure}[t]
\begin{tabular}{ccc}
\begin{minipage}{.33\textwidth}
\centering
\caption*{\small $\varepsilon=0.03$}
\vspace{-0.3cm}
\includegraphics[width=0.90\linewidth]{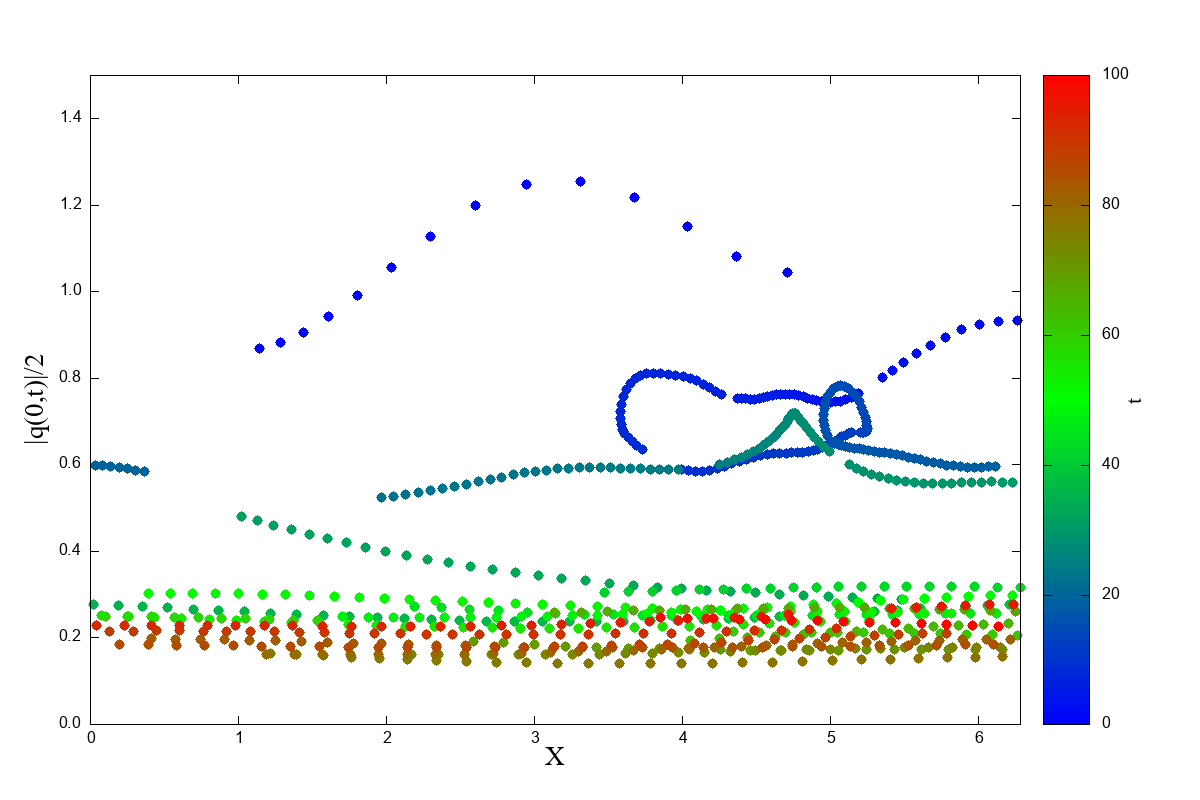}
\end{minipage}
\begin{minipage}{.33\textwidth}
\centering
\caption*{\small $\varepsilon=0.04$}
\vspace{-0.3cm}
\includegraphics[width=0.90\linewidth]{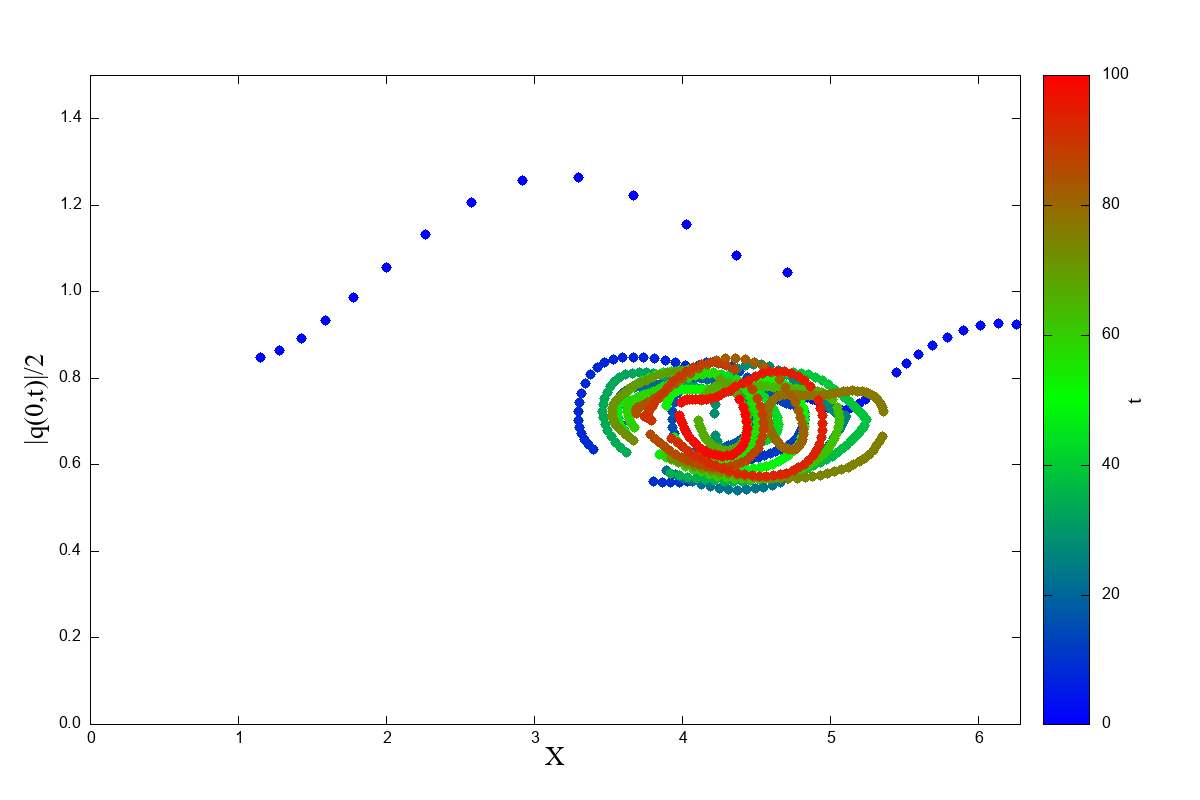}
\end{minipage} 
\begin{minipage}{.33\textwidth}
\centering
\caption*{\small $\varepsilon=0.05$}
\vspace{-0.3cm}
\includegraphics[width=0.90\linewidth]{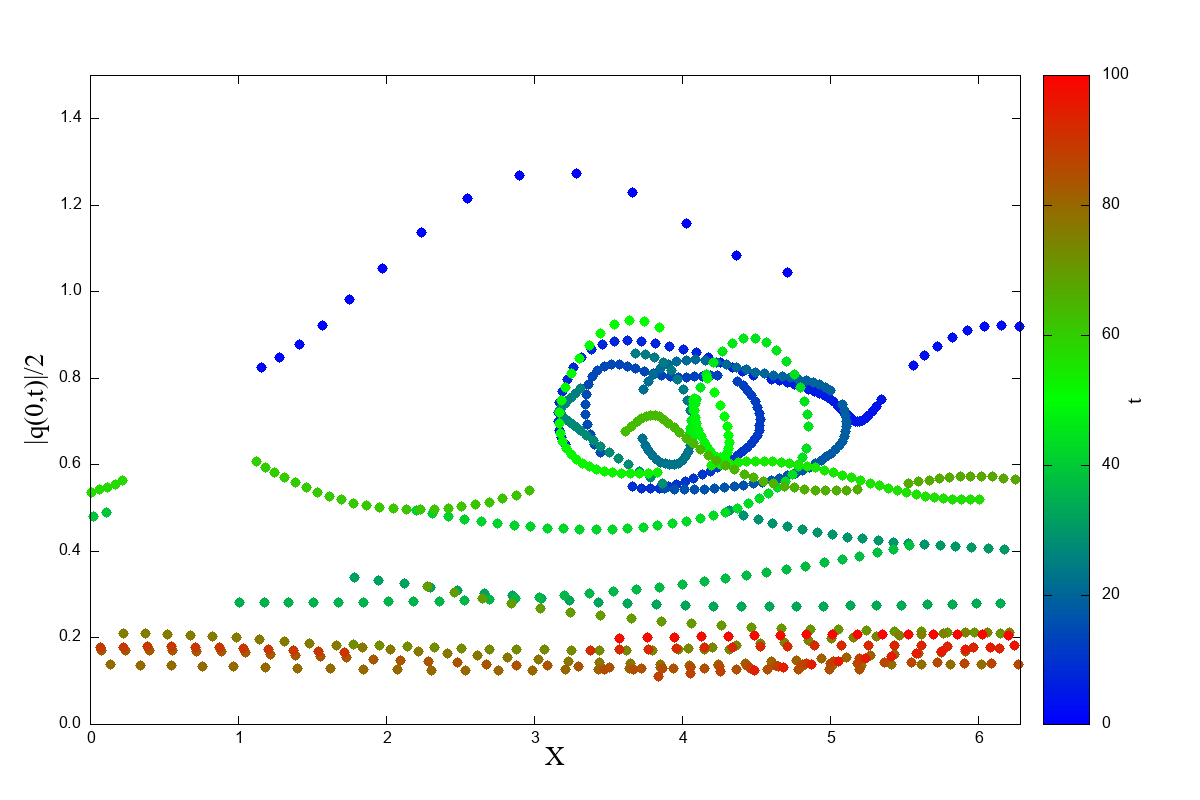}
\end{minipage}
\end{tabular}
\caption{\label{fig:attractor_comparison_sec2}
The trajectories with $\varepsilon=0.03$(left), 0.04(middle), 0.05(right).}
\end{figure}

Since the chaotic behavior is quite sensitive to the initial condition, 
even a slight variation of the perturbation parameters with a fixed initial condition might result 
in significant changes later on.
Here we present the results for several strength of the forcing terms $\varepsilon(=\varepsilon_1=\varepsilon_2)$.
We fix the initial condition $A=1.2,\eta_0=0.871$ and the coefficient of the dissipative term to $\gamma=0.05$. 
The results are shown in Fig.~\ref{fig:attractor_comparison_sec2}, where the 
attractor is observed for $\varepsilon=0.04$. On the other hand, 
the trajectories collapse as increasing time in $\varepsilon=0.03$ and $\varepsilon=0.05$.
Therefore, the Bekki-Nozaki equation's chaotic nature depends on the perturbation parameters.

\begin{figure}[t]

\includegraphics[width=1\linewidth]{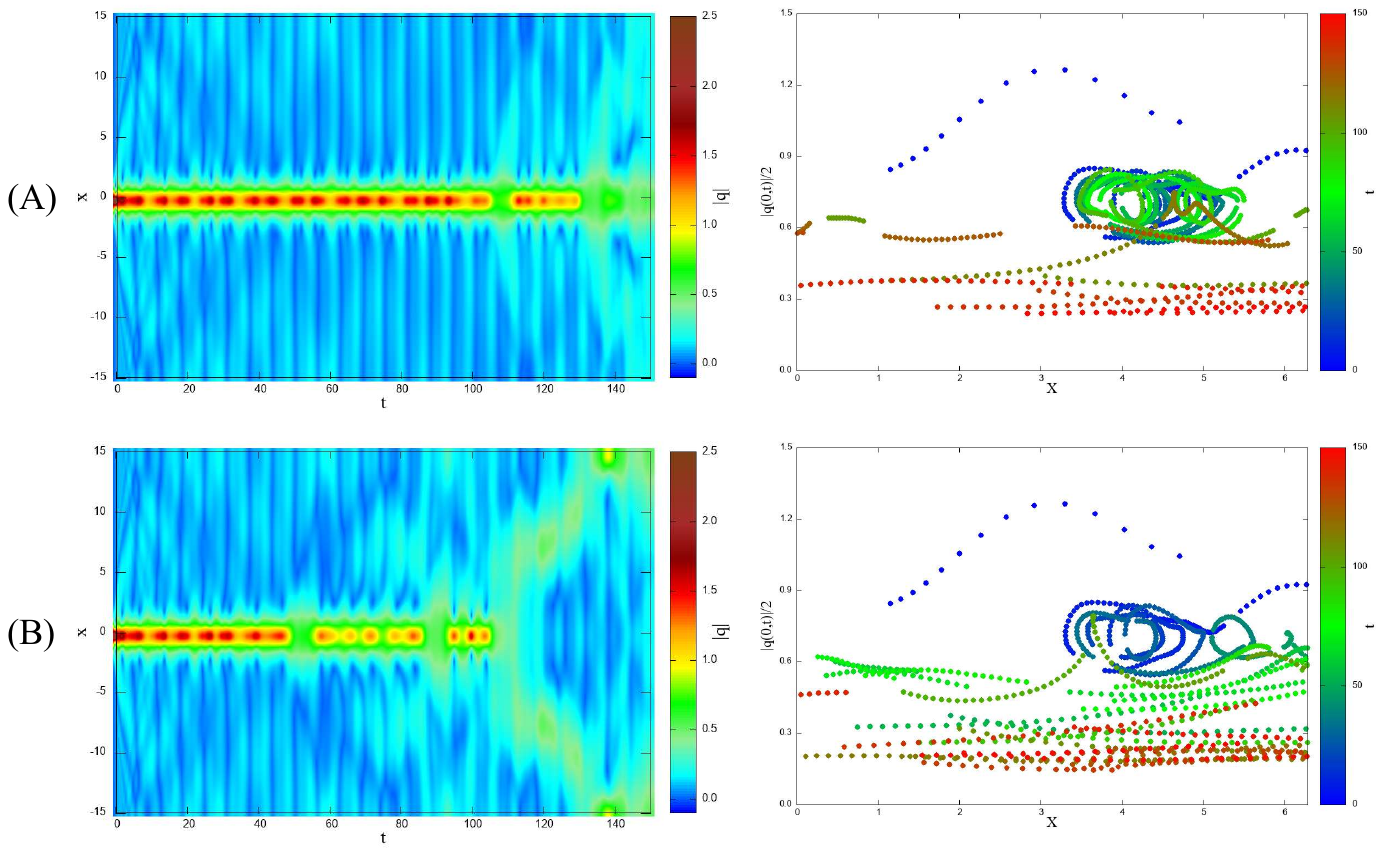}
\vspace{-2.0cm}
\caption{\label{fig:dt_sec2}
The solutions of Eq.~\eqref{eq:Bekki-Nozaki_sec1} computed 
by (A) $dt=10^{-4}$, (B) $dt=10^{-5}$: the contour plot (left), the trajectory (right).}
\end{figure}

Once the parameters are fixed, however, changing the conditions of the numerical analysis 
may provide quite different results.
We show the result of varying the integration time step with the same parameters set as in previous studies:
$A=1.2,\eta_0=0.871,\varepsilon=0.0405,\gamma=0.05$, and $\omega=1.0$.
In Fig.\ref{fig:dt_sec2}, we present the contour plot and the trajectory of the solutions computed with
$dt=10^{-4}$ and $dt=10^{-5}$. Because the norm of the solutions decays and the attractor collapses 
for more than 150-seconds, we plot the behavior of the solutions over a longer time period. 
This information will be useful for our study. 
The different $dt$ alter the behavior of the solutions, even when the initial condition and perturbative terms 
remain completely the same. It is worthwhile to examine the chaos system using a methodology which does not 
explicitly depend on the 
computational mesh or the integration scheme.
In this paper, we employ a novel numerical method called Physics-Informed Neural Networks (PINNs), 
to solve PDEs without assuming any discretizing scheme. We briefly review PINNs in the next section.  

\section{Physics-Informed Neural Networks\label{sec:3pinns}}

Recently, a deep learning method based on artificial neural networks called Physics-Informed Neural Networks (PINNs) 
has grown much attention as a novel approach for solving several nonlinear PDEs.
PINNs integrate physical laws into the learning process, yielding solutions that closely align with the underlying physics.
PINNs can solve both forward problems, where approximate solutions to governing equations are obtained, and inverse problems, 
where the coefficients of the equations are determined using training data.

\subsection{Architecture}

\begin{figure}[b]
\centering
\includegraphics[width=23pc]{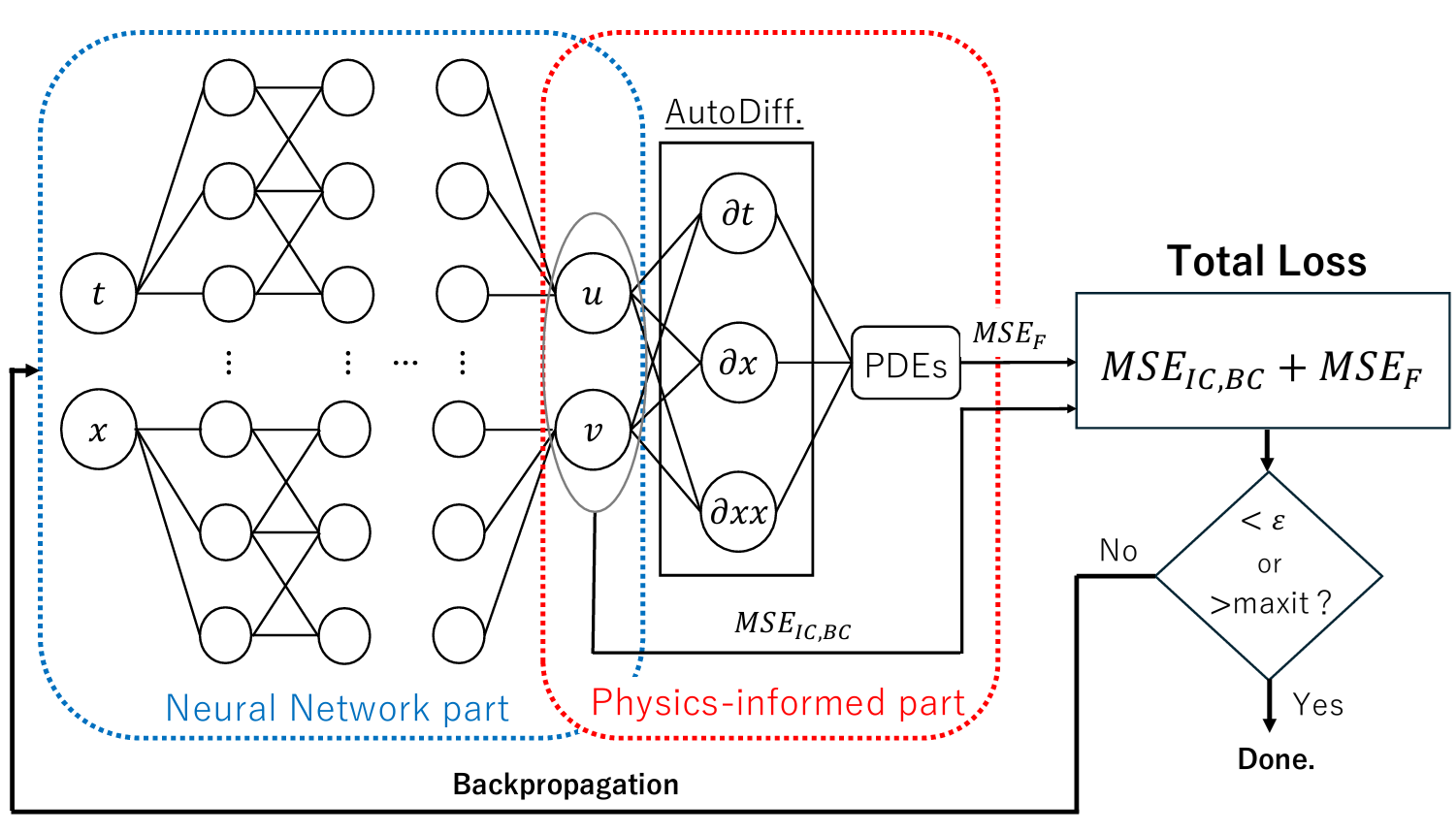}
\caption{\label{fig:pinns_sec3}
The whole architecture of PINNs: The neural network part consists of input layers, hidden layers, and output layers.
In the physics-informed part, the derivatives of the output functions are calculated using automatic differentiation and the PDEs are defined.}
\end{figure}

Figure~\ref{fig:pinns_sec3} shows schematic diagram of PINNs.
The neural network part is made up of input layers, hidden layers, and output layers, as traditional neural networks.
Outputs of the first part, i.e. the functions ($u,v$) are used to evaluate the derivatives in terms of the automatic differentiation~\cite{automaticdifferentiation} in the second part.
This is so called the physics-informed part.
Then, the loss function is formed, which is composed of the PDEs with the derivatives, the boundary conditions, and the initial conditions.

\subsection{Forward Analysis}

PINNs obtain the solutions to the PDEs by optimizing neural networks to approximate functions that satisfy the initial and the boundary conditions.
The solution of \eqref{eq:Bekki-Nozaki_sec1} is a complex-valued function and it is efficient to introduce real functions $u,v$ as $q(x,t):=u(x,t)+iv(x,t)$.
Then, we consider two PDEs corresponding to \eqref{eq:Bekki-Nozaki_sec1}
\begin{align}
F_\mathrm{real} &:= u_t + \mathcal{N}_\mathrm{real}(u,v,u_t,v_t,u_x,v_x,u_{xx},v_{xx},\cdots) = 0,
\\
F_\mathrm{imag} &:= v_t + \mathcal{N}_\mathrm{imag}(u,v,u_t,v_t,u_x,v_x,u_{xx},v_{xx},\cdots) = 0,
\label{eq:pde_forward_sec3}
\end{align}
where
\begin{align}
\mathcal{N}_\mathrm{real} &= v_{xx} + 2(u^2 + v^2)v
-\varepsilon_1 \cos(\omega t) -\varepsilon_2 \cos(2\omega t)
-\gamma u_{xx},\\
\mathcal{N}_\mathrm{imag} &= u_{xx} + 2(u^2 + v^2)u
-\varepsilon_1 \sin(\omega t) -\varepsilon_2 \sin(2\omega t)
-\gamma v_{xx}.
\end{align}
To optimize the networks, we define a loss function as the sum of the mean squared errors (MSEs) for the initial/boundary conditions and the PDEs.
The loss function is expressed as follows:
\begin{align}
Loss_\mathrm{forward}
&= \frac{1}{N_u} \sum_{N_u}|IC|^2
+ \frac{1}{N_b} \sum_{N_b}|BC|^2 
+ \frac{1}{N_F} \sum_{N_F}
\left(|F_{real}|^2 + |F_{imag}|^2\right).
\label{eq:loss_forward_sec3}
\end{align}
The first term represents the loss concerning the initial conditions.
It measures the discrepancy between the neural network's predicted values $u_{\mathrm{pred}}^0,v_{\mathrm{pred}}^0$ and the training data $u_{\mathrm{correct}}^0,v_{\mathrm{correct}}^0$ as follows:
\begin{align}
\frac{1}{N_u} \sum_{N_u}|IC|^2
= &\frac{1}{N_u}\sum_{i=1}^{N_u}|u_{\mathrm{pred}}^{0}(x^i,0)
- u_\mathrm{correct}^{0}(x^i,0)|^2
\notag\\
+ &\frac{1}{N_u}\sum_{i=1}^{N_u}|v_{\mathrm{pred}}^{0}(x^i,0)
- v_\mathrm{correct}^{0}(x^i,0)|^2.
\label{eq:loss_forward_u_sec3}
\end{align}
Here, $N_u$ is the number of reference points in the training data used for learning.
The second term represents the loss concerning the boundary conditions with collocation points $N_b$.
These are usually referred to as the data loss, which is directly evaluated for the estimated functions in terms of the neural networks.
The third term is the loss of the PDEs, which is called physical loss.
It supports the estimated functions satisfying the physical laws, expressed as the PDEs.
PINNs optimize the network parameters to minimize the $Loss$~\eqref{eq:loss_forward_sec3}.
As a result, the estimated functions that satisfy both the initial/boundary conditions and the PDEs are obtained when $Loss\rightarrow0$.

\subsection{Inverse Analysis}

Another significant application of PINNs is the inference of PDEs from training data.
We now consider parameterized PDEs
\begin{align}
\tilde{F}_\mathrm{real} &:= u_t + 
\tilde{\mathcal{N}}_\mathrm{real}(u,v,u_t,v_t,u_x,v_x,u_{xx},v_{xx},\cdots;\bm{\lambda}) = 0
\label{eq:pde1_inverse_sec3},
\\
\tilde{F}_\mathrm{imag} &:= v_t + 
\tilde{\mathcal{N}}_\mathrm{imag}(u,v,u_t,v_t,u_x,v_x,u_{xx},v_{xx},\cdots;\bm{\lambda}) = 0,
\label{eq:pde2_inverse_sec3}
\end{align}
where $\bm{\lambda}$ represents the parameters of the PDEs.
Solving inverse problems means accurately estimating the parameters $\bm{\lambda}$.
To solve these problems, the loss function is constructed as follows:
\begin{align}
Loss_\mathrm{inverse}
&= \frac{1}{N_u} \sum_{N_u}
\left(|u_{\mathrm{pred}} - u_\mathrm{correct}|^2
+ |v_{\mathrm{pred}} - v_\mathrm{correct}|^2\right)
\notag \\
&+ \frac{1}{N_F} \sum_{N_F}
\left(|\tilde{F}_\mathrm{real}|^2 + |\tilde{F}_\mathrm{imag}|^2\right).
\label{eq:loss_inverse_sec3}
\end{align}
The first term corresponds to the training data, while the second term enforces the structure imposed by Eqs.~\eqref{eq:pde1_inverse_sec3} and \eqref{eq:pde2_inverse_sec3} at a finite set of collocation points, $N_u$ and $N_f$.
The parameters $\bm{\lambda}$ can be determined by minimizing \eqref{eq:loss_inverse_sec3}.

\subsection{Numerical Setup}

We employ the Adam optimizer and also  
the limited-memory Broyden-Fletcher-Goldfarb-Shanno (L-BFGS) optimizer~\cite{adam,l-bfgs} for minimizing the loss functions.
The former is a simple stochastic gradient descent (SGD) algorithm, 
and the latter is a gradient descent optimization algorithm using a quasi-Newton method.
The efficiency and the convergence speed are certainly increased by combining these optimization algorithms. 
During the early stages of training, the Adam is used to efficiently progress the learning process and 
subsequently the L-BFGS allows for fast and high-precision optimization.
For the activation function, the hyperbolic tangent function is used.
Throughout the analysis in this paper, VScode is used as the Integrated Development Environment for Python 3.12, 
and TensorFlow 2.18.0 is used for building PINNs.
Adam from TensorFlow and L-BFGS from Scipy 1.15.1. are used for optimization.
The computations are performed on a 64-bit Windows system with an Intel(R) Core(TM) i9-14700KF CPU and NVIDIA 
GeForce RTX 4090 GPUs.

\section{PINNs analysis for the Bekki-Nozaki equation\label{sec:4analyses}}

As discussed in Section~\ref{sec:2bekkinozaki}, the chaotic behavior of the solution to Eq.~\eqref{eq:Bekki-Nozaki_sec1} can be achieved by the right initial conditions and perturbative parameters.
Note that it also depends on the setup of numerical analysis, such as the computational mesh and the integration scheme.
It is worth examining the PDEs~\eqref{eq:Bekki-Nozaki_sec1} by using PINNs since 
they may demonstrate that the chaos of the equation, not the numerical artifact, has inherently chaotic properties. 
This highlights the importance of using PINNs to analyze several PDEs.

For plotting a trajectory of the attractor in the phase space $(\chi(t),|q(x_m,t)|/2)$, 
the solutions over a sufficiently long-time interval are required.
Although PINNs can formally treat the entire time domain at once without any discretization, 
the computational cost for long-term simulations with high accuracy becomes so expensive.
To address the issue, we employ a method of dividing the time interval into small segments and solve the problems sequentially.
Specifically, for the 30 second solution, we divide the time interval into the blocks such as 
$t\in[0,2]$, $t\in[2,4]$, $t\in[4,6]$, and so on, solving each segment step by step.
Use this prescription, we are able to effectively balance computational efficiency with long-term prediction accuracy.

We start with the result of the forward-PINNs analysis.
The network architecture consists of 4 hidden layers, each with 128 nodes.
The number of training points is set as $N_u = 1,000$, $N_b = 2,000$, and $N_F = 50,000$.
We show the contour plot of the PINNs solution over the time interval $t \in [0,30]$ in Fig.6(a) and the plot of the attractor in Fig.6(b).
\begin{figure}[t]
\begin{tabular}{cc}
\begin{minipage}{.48\textwidth}
\centering
\includegraphics[width=1.10\linewidth]{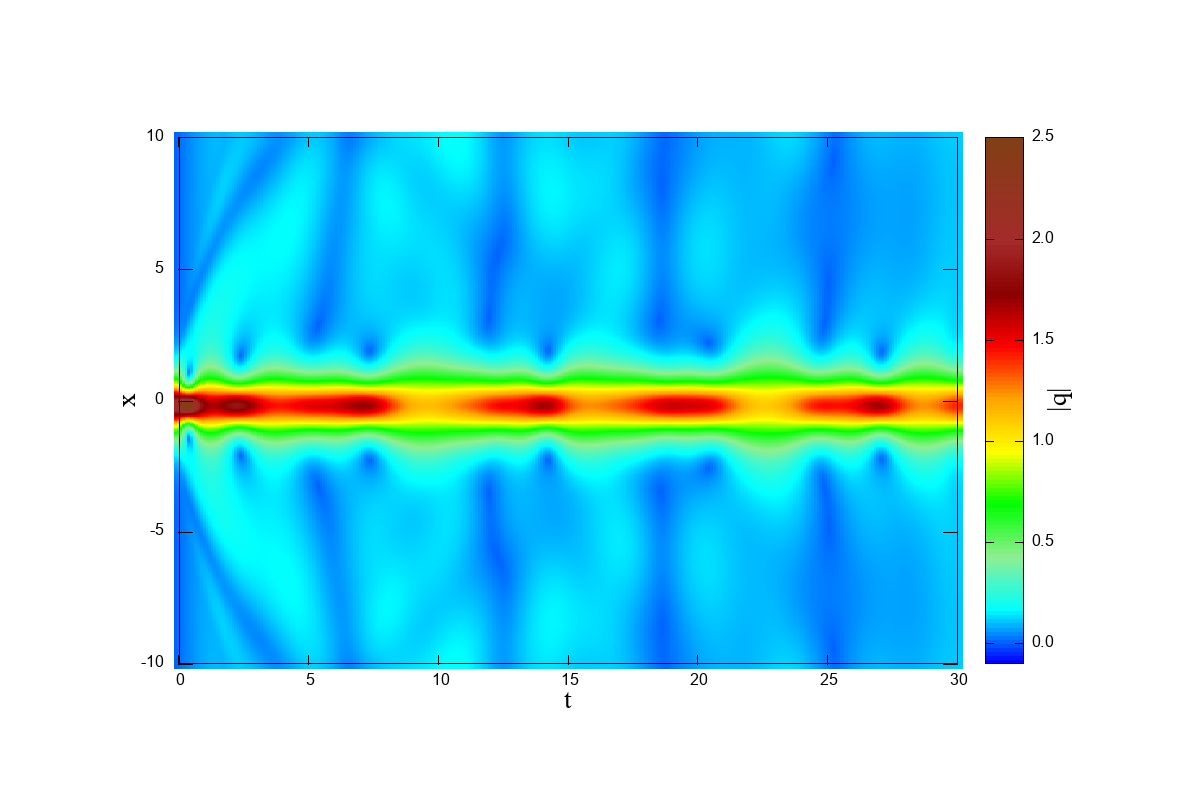}
\subcaption{\label{fig:contour_pinns_sec4}
The contour plot}
\end{minipage} 
\begin{minipage}{.48\textwidth}
\centering
\vspace{0.5cm}
\includegraphics[width=0.90\linewidth]{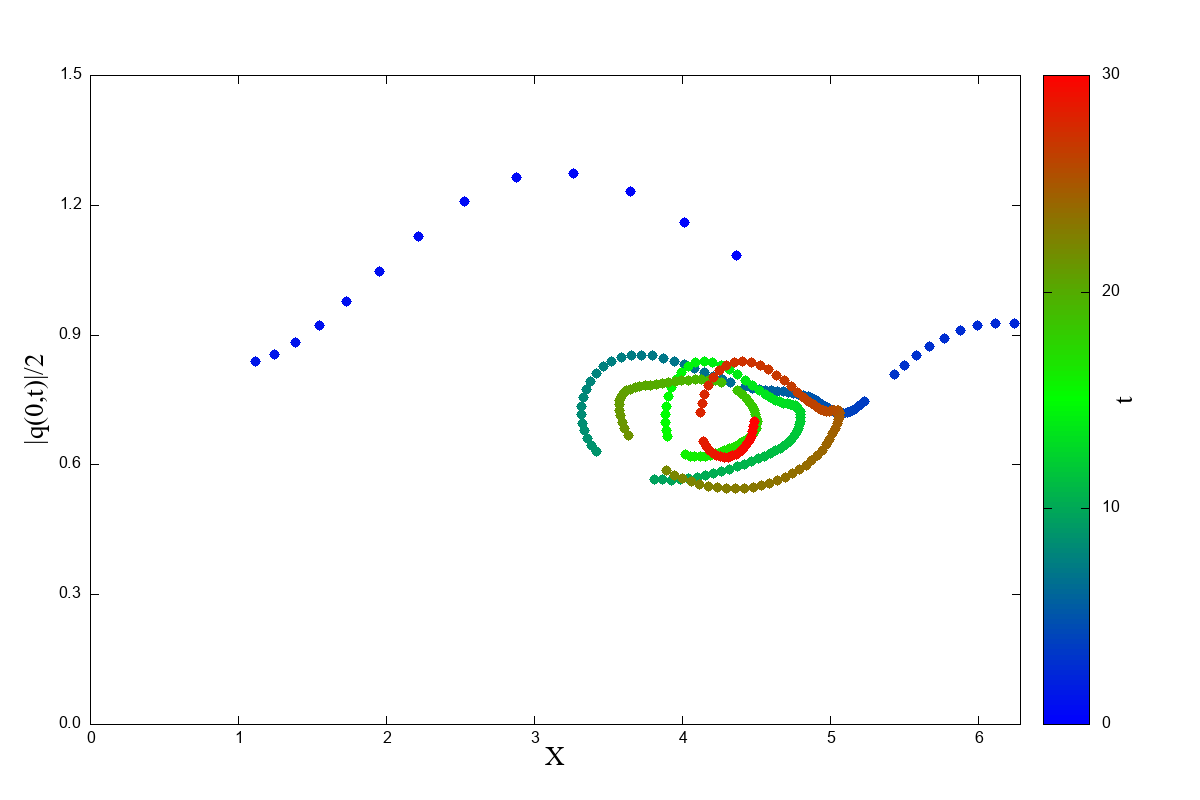}
\vspace{0.5cm}
\subcaption{\label{fig:attractor_pinns_sec4}
The trajectory}
\end{minipage}
\end{tabular}
\caption{\label{fig:solution_pinns_sec4}
The behavior of the Forward-PINNs solution.}
\end{figure}
As you see that the forward-PINNs solution also exhibits the bouncing of norm and the presence of the attractor.
The solution of PINNs, the mesh-free algorithm, also has a property of chaos.
This implies that the chaos in the Bekki-Nozaki equation is not a numerical artifact, but rather an inherent property of the system itself.

Next, we show the results of the inverse-PINNs analysis.
The forward-PINNs solution for $t \in [0,30]$ is used as training data for the estimation of the perturbation parameters through the 
inverse-PINNs.
In this case, we set the hidden layers to 6 layers with 40 nodes each, and the number of training points to $N_u = N_F = 300,000$.
Since the training data is randomly extracted, we do the analysis in a few times and compare the output. 
The results are summarized in Table~\ref{tab:i-pinns-1_sec4}, which
indicate that PINNs accurately identify the perturbative parameters.
Since the solution obtained by forward-PINNs is successfully identified as the solution to the Bekki-Nozaki equation, 
we can conclude that the forward analysis performs correctly.
Furthermore, this result demonstrates that inverse-PINNs analysis works well even with chaotic data.
\begin{table}[b]
\centering
\caption{\label{tab:i-pinns-1_sec4}
The results of the identification of the perturbation parameters}
\begin{tabular}{|c|c|c|c|c|} \hline
  &$\varepsilon_1$&$\varepsilon_2$&$\gamma$&$Loss$\\
\hline
Correct&0.0405&0.0405&0.05& \\ 
Identified(1)&0.04046&0.04032&0.05039&$1.84\times10^{-5}$  \\
Identified(2)&0.04067&0.04019&0.05122&$3.12\times10^{-5}$  \\
Identified(3)&0.04052&0.04039&0.05101&$1.52\times10^{-5}$  \\
\hline
\end{tabular}
\end{table}

\textcolor{red}{}

\begin{figure}[t]
\begin{tabular}{cc}
\begin{minipage}{.48\textwidth}
\centering
\includegraphics[width=1.0\linewidth]{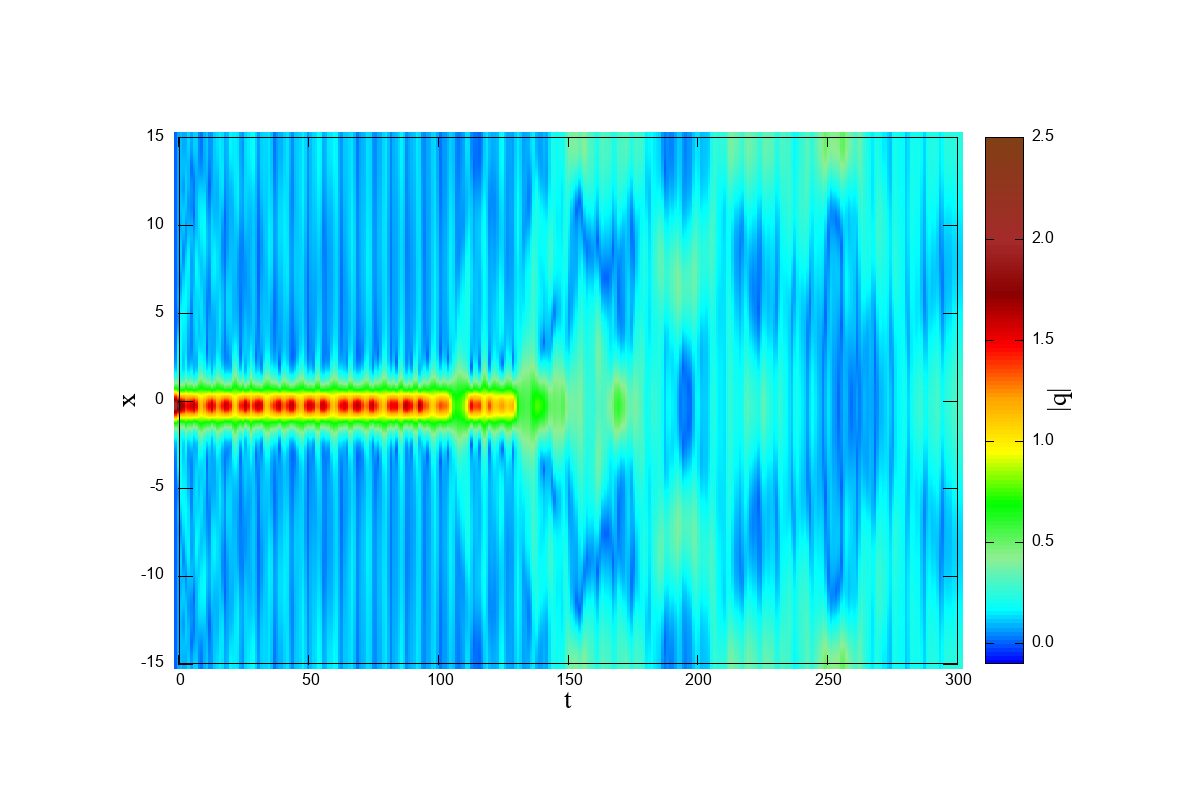}
\subcaption{\label{fig:contour_rk_A_sec4}
The parameter set (A)}
\end{minipage}
\begin{minipage}{.48\textwidth}
\centering
\includegraphics[width=1.0\linewidth]{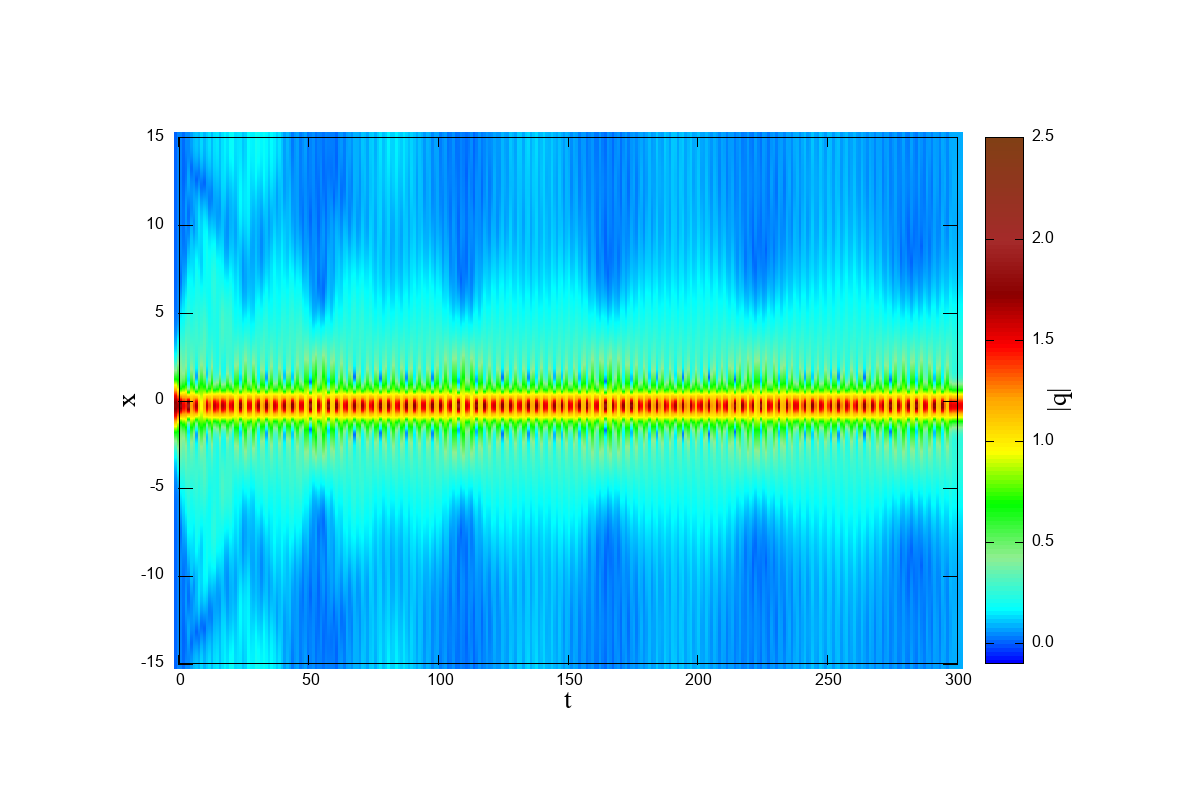}
\subcaption{\label{fig:contour_rk_B_sec4}
The parameter set (B)}
\end{minipage}
\end{tabular}
\caption{\label{fig:solution_rk_sec4}
The solutions of Runge-Kutta method for parameter set (A) and (B).}
\end{figure}

Finaly, we study the dynamical transition of the solution; the decay of the chaotic phase into an oscillating vacuum.
As shown in Fig.~\ref{fig:dt_sec2}, the solution decays over a longer time scale.
It is worth investigating the use of inverse-PINNs to analyze the transition of solutions into the non-chaotic ones 
and how the coefficients' identification changes.
We investigate the two parameter sets that exhibit different chaotic behavior.
The parameter set (A) is $\varepsilon_1=\varepsilon_2=0.0405, \gamma=0.05$, and $\omega=1.0$ 
and $A=1.2,\eta_0=0.871$. 
The parameter set (B) is $\varepsilon_1=0,\varepsilon_2=0.0405,\gamma=0.05$, 
and $\omega=1.0$, and $A=1.7,\eta_0=0.5167$.
The behavior of the solutions for these parameters are shown in Fig.~\ref{fig:solution_rk_sec4}.
The solution of (A) dissipates at about $t = 150$ second, but the solution of (B) keeps bouncing for a long time.

\begin{figure}[t]
\begin{minipage}{0.8\textwidth}
\centering
\includegraphics[width=1.00\linewidth]{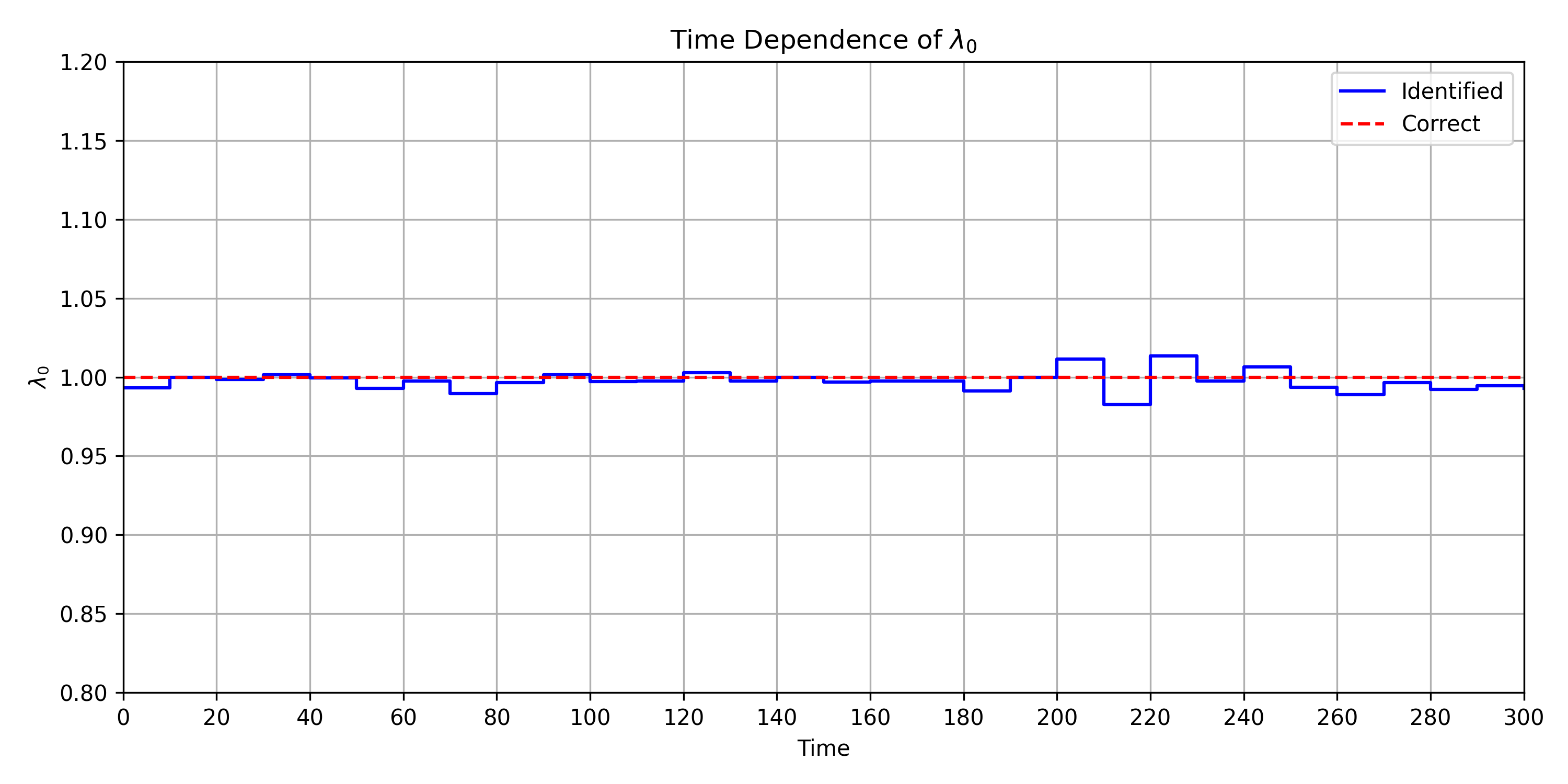}
\subcaption{\label{fig:i-pinns_a_A12-eta0871_dt10_sec4}
The dispersion coefficient $\lambda_0$}
\end{minipage} 
\begin{minipage}{0.8\textwidth}
\centering
\includegraphics[width=1.00\linewidth]{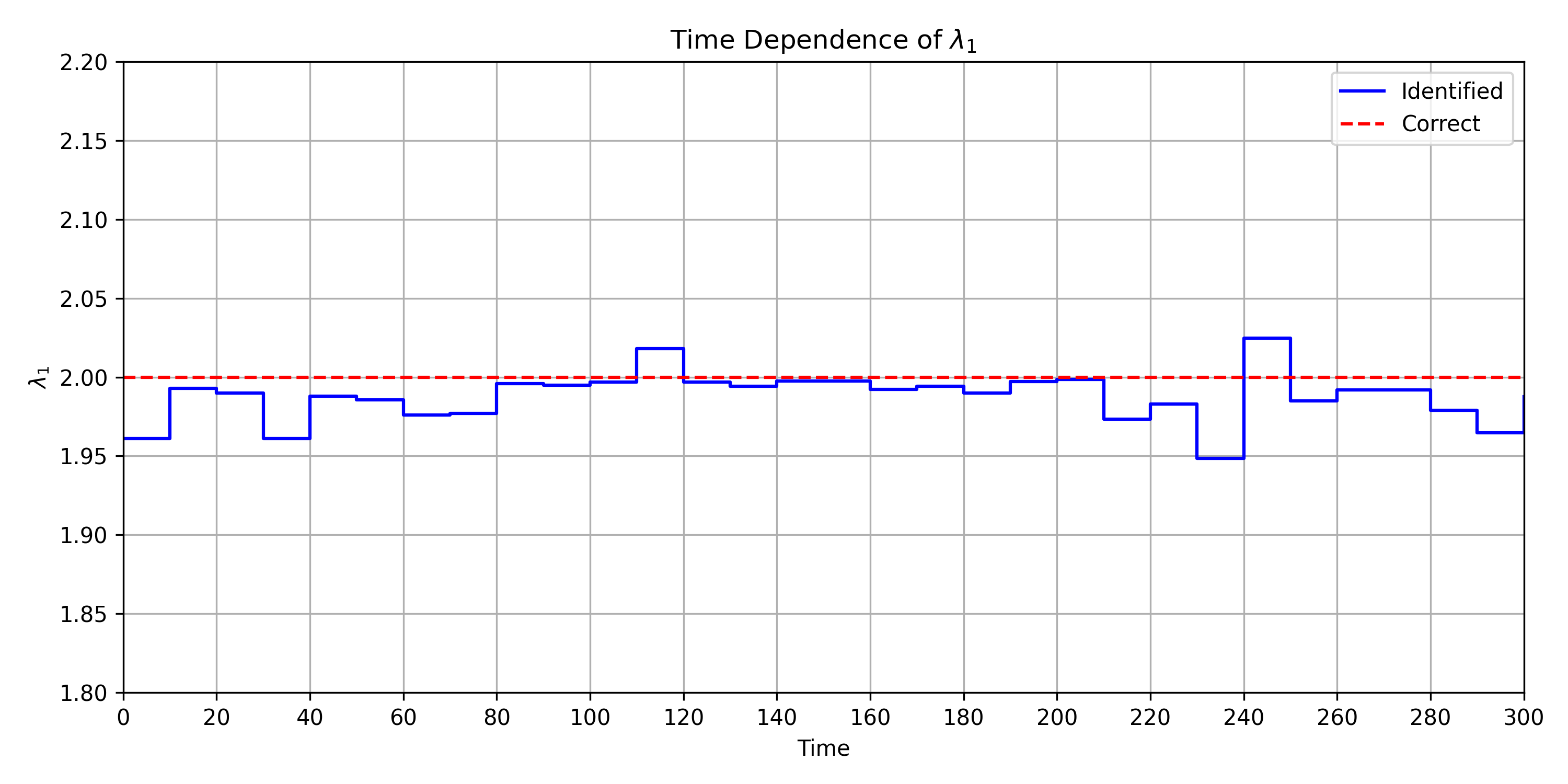}
\subcaption{\label{fig:i-pinns_b_A12-eta0871_dt10_sec4}
The nonlinear coefficient $\lambda_1$}
\end{minipage}
\caption{\label{fig:i-pinns_integrable_A12-eta0871_dt10_sec4}
The results of the identification for the coefficients in integrable NLS eq with the 10-second segment.}
\end{figure}

\begin{figure}[t]
\begin{minipage}{0.8\textwidth}
\centering
\includegraphics[width=1.00\linewidth]{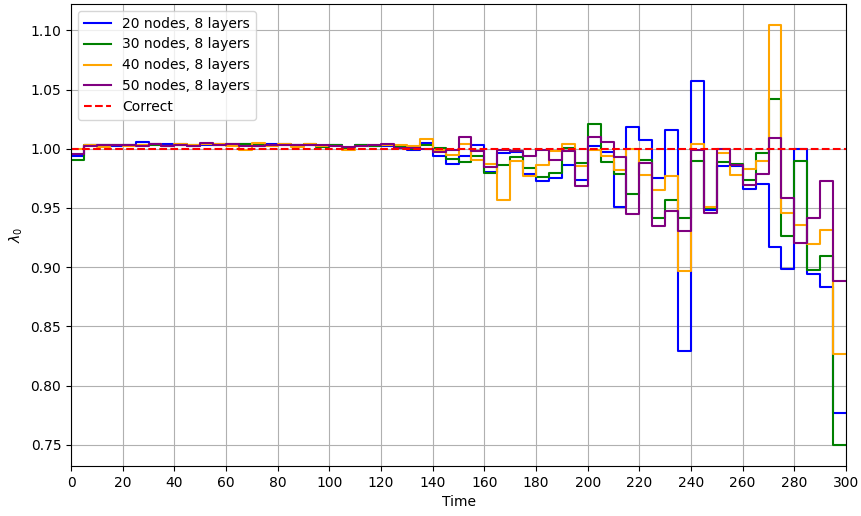}
\subcaption{\label{fig:i-pinns_a_A12-eta0871_dt5_sec4}
The dispersion coefficient $\lambda_0$}
\end{minipage} 
\begin{minipage}{0.8\textwidth}
\centering
\includegraphics[width=1.00\linewidth]{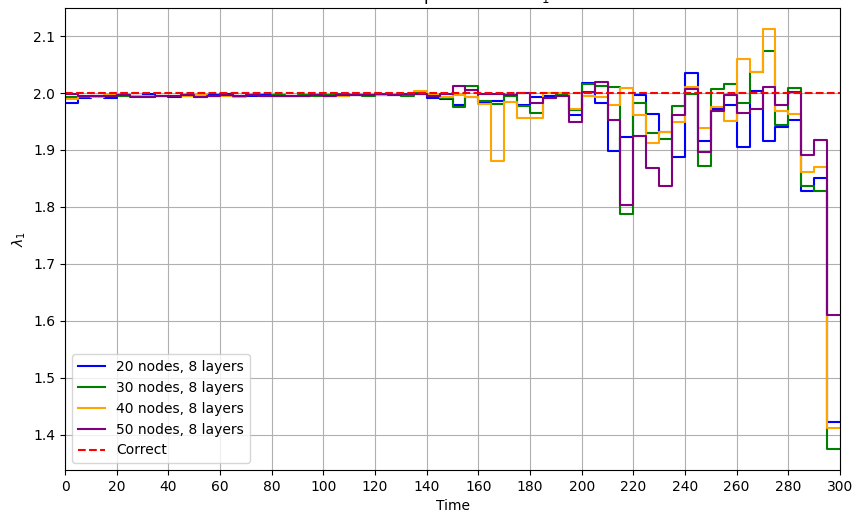}
\subcaption{\label{fig:i-pinns_b_A12-eta0871_dt5_sec4}
The nonlinear coefficient $\lambda_1$}
\end{minipage}
\caption{\label{fig:i-pinns_integrable_A12-eta0871_dt5_sec4}
The results of the identification for the coefficients in integrable NLS eq. of parameter set (A) with the 5-second segment. }
\end{figure}

\begin{figure}[t]
\begin{minipage}{0.8\textwidth}
\centering
\includegraphics[width=1.00\linewidth]{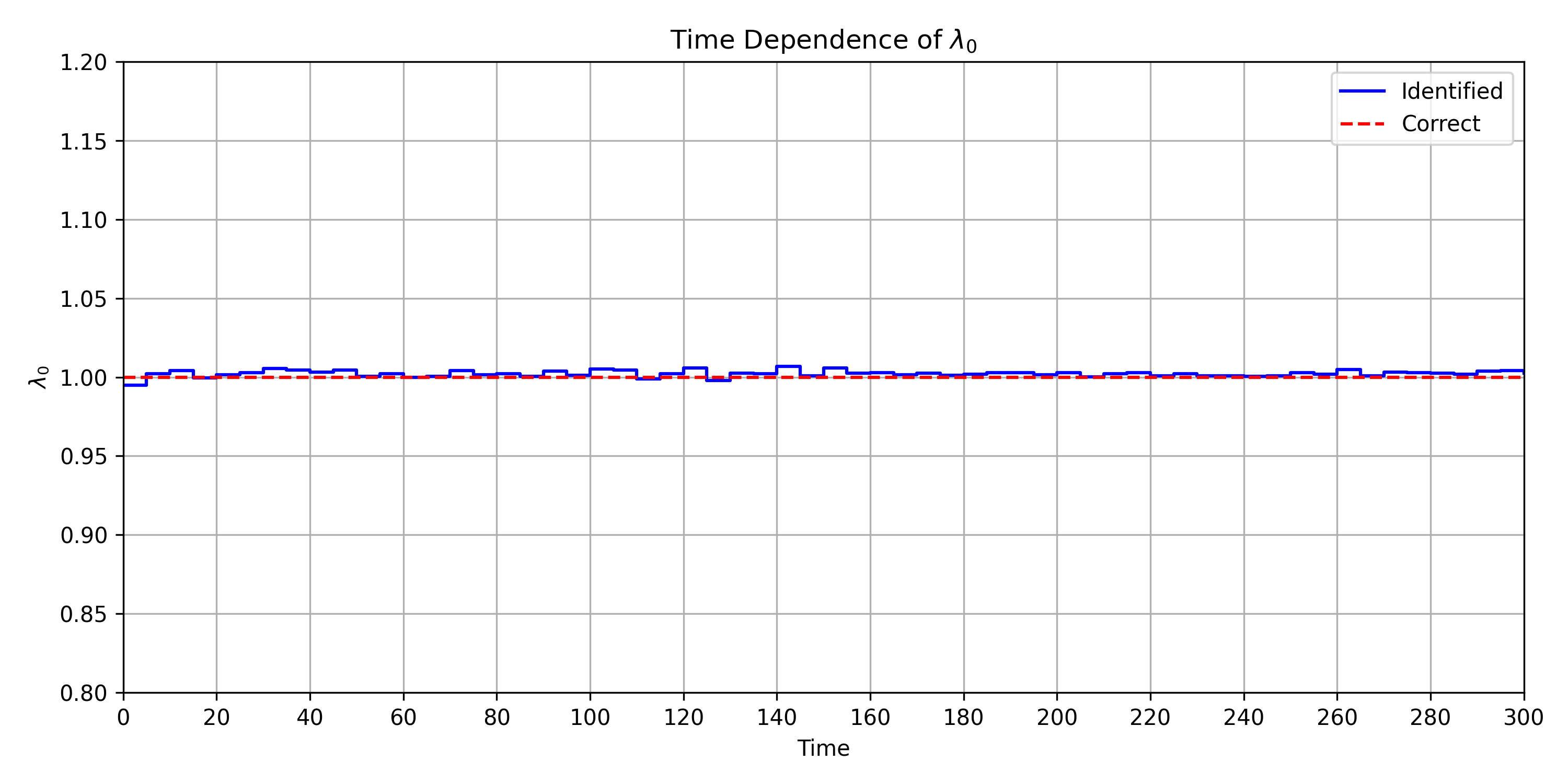}
\subcaption{\label{fig:i-pinns_a_A17-eta05167_dt5_sec4}
The dispersion coefficient $\lambda_0$}
\end{minipage} 
\begin{minipage}{0.8\textwidth}
\centering
\includegraphics[width=1.00\linewidth]{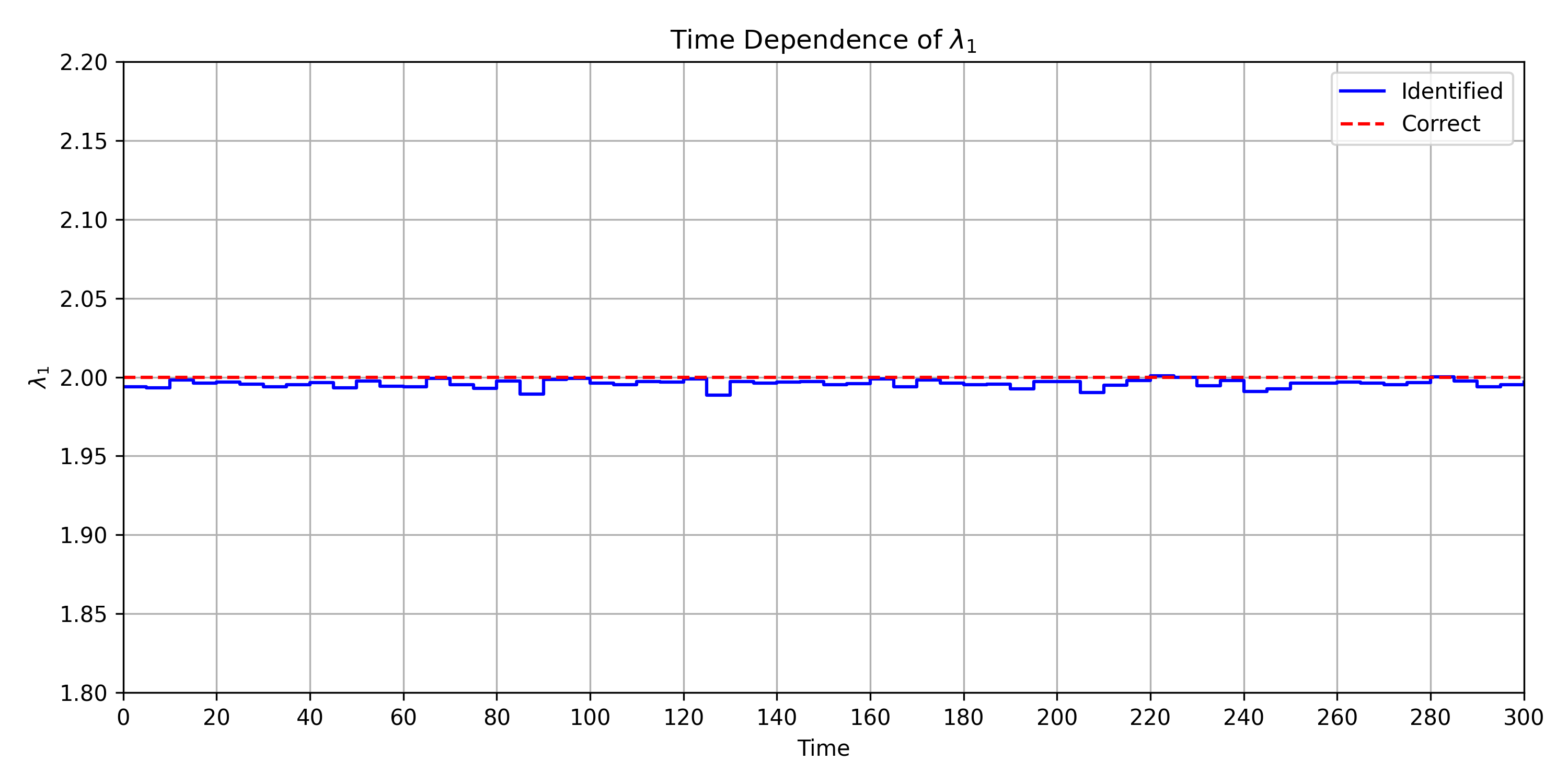}
\subcaption{\label{fig:i-pinns_b_A17-eta05167_dt5_sec4}
The nonlinear coefficient $\lambda_1$}
\end{minipage}
\caption{\label{fig:i-pinns_integrable_A17-eta05167_dt5_sec4}
The results of the identification for the coefficients in integrable NLS eq. of parameter set (B) with the 5-second segment.}
\end{figure}

In the inverse analysis, 
we fix the coefficients of the perturbation terms and estimate the dispersion and non-linear coefficients 
of the original NLS equation. We define the governing equation for the inverse PINNs as
\begin{align}
iq_t + \lambda_0 q_{xx} + \lambda_1 |q|^2q 
= 0.0405i \left\{\exp(it) + \exp(2it)\right\} + 0.05i q_{xx}
\label{eq:i-pinns_A_integrable_sec4}
\end{align}
of the set (A) and
\begin{align}
iq_t + \lambda_0 q_{xx} + \lambda_1 |q|^2q 
= 0.0405i \exp(2it) + 0.05i q_{xx}
\label{eq:i-pinns_B_integrable_sec4}
\end{align}
of the set (B).
The exact values of the parameters are $\lambda_0=1.0$ and $\lambda_1=2.0$.
Analyzing the long-term behavior and examine whether the structure of the PDE changes 
in response to the solution's behavior, we do not train the whole data set at once.
Instead, we divide the time domain $t \in [0, 300]$ into smaller segments.
In this study, we perform the analysis using the 10-second and the 5-second segments. 
The number of training points is set to $N_u = N_F = 100{,}000$ for the 10-second segments, and $50{,}000$ for the 5-second segments.
Fig.~\ref{fig:i-pinns_integrable_A12-eta0871_dt10_sec4} presents the result of estimating the coefficients of the dispersion and non-linear terms with the 10-second segment for parameter set (A), as in \eqref{eq:i-pinns_A_integrable_sec4}.
Although there are small discrepancies, the coefficients are generally accurately estimated both before and after the solution dissipates.
On the other hand, with the 5-second segment, the discrepancy in the coefficients begins to fluctuate 
when the solution's norm decays and the attractor collapsed as shown in Fig.~\ref{fig:i-pinns_integrable_A12-eta0871_dt5_sec4}.
We perform the analysis with changing the number of nodes (20, 30, 40, 50) in each of the 8 hidden layers, and find that the results are similar regardless of the number of nodes.
For the case of parameter set (B), the dispersion and non-linear coefficients are correctly estimated with 5-second segment.
Fig.~\ref{fig:i-pinns_integrable_A17-eta05167_dt5_sec4} shows the result of the estimation.
Based on this, we can conclude that PINNs provide accurate estimates of coefficients regardless of the size of the time segment so long as the solution continues to bounce.

When the solution continues the bouncing, the NLS soliton works as an attractor of the chaos and then, the dynamics of the system 
are mainly governed by the original non-linear Schr\"{o}dinger equation.
On the other hand, after the soliton start to decay, the external forcing becomes dominate the system's dynamics.
Since the period of the external forcing is $2\pi$, it can be inferred that if the time segment is longer than the 
forcing period, the system’s behavior can be accurately captured.
However, if the segment is shorter than the period, it fails to correctly interpret the system's dynamics.

\section{Summary}

In this paper, we investigated the PINNs to analyze the chaos in the non-linear Schr\"{o}dinger equation
with forced dissipative terms, known as the Bekki-Nozaki equation.
The traditional numerical methods reveal that chaotic behavior inevitably might be caused from the computational mesh.
By using PINNs which are free from choice of the mesh, we showed that the Bekki-Nozaki equation inherently exhibits chaotic behavior.
Furthermore, when the system becomes non-chaotic over time, the inverse-PINNs analysis fails to estimate the NLS equation's 
dispersion and nonlinear terms correctly.
This is concluded to be a result of the system's behavior shifting away from chaos with a soliton solution as the attractor, 
towards a non-chaotic state.

\vspace{0.5cm}

\section*{References}

\bibliography{references}

\end{document}